\documentstyle[aps,prb]{revtex}
\begin{document}

\twocolumn

\title{Optical properties of arrays of quantum dots with internal
disorder}

\author{E. V. Tsiper}
\address{Department of Physics, University of Utah, Salt Lake City,
Utah 84112$^*$}

\maketitle

\begin{abstract}
Optical properties of large arrays of isolated quantum dots are
discussed in order to interpret the existent photoluminescence
data. The presented theory explains the large observed shift between
the lowest emission and absorption energies as the average distance
between the ground and first excited states of the dots.
The lineshape of the spectra is calculated for the case when the
fluctuations of the energy levels in quantum dots are due to the
alloy composition fluctuations. The calculated lineshape is in good
agreement with the experimental data. The influence of fluctuations
of the shape of quantum dots on the photoluminescence spectra is also
discussed.
\end{abstract}


\section{Introduction}
Reduced-dimensionality structures are currently attracting much
attention.\cite{P5,M1,K2,P6} Modern technology made it possible to
create nanoscale 0D structures where motion in all directions is
quantized (quantum dots). Often a system represents a large array of
independent quantum dots. Because of extremely small dimensions,
the fluctuations of parameters of individual quantum dots become an
important factor since they determine the properties of the whole
array.

This paper was initiated by the experimental work Ref.\
\onlinecite{P5}, where nice photoluminescence (PL) and
photoluminescence excitation (PLE) experiments were performed with an
array of self-assembling quantum dots. The conventional nonselective
(with above-barrier excitation) PL reveals a broad peak of about
$\sim$ 50 meV in width. This width originates from a wide spread of
energies of different quantum dots in the array. Both PLE and
selectively-excited PL spectra (when only the quantum dots that are
in resonance with incident light are excited) show sets of broadened
peaks with 2-3 times smaller widths. These peaks correspond to the
distribution of energy levels in the subset of quantum dots that are
resonantly excited. It was observed that the first PLE peak is
strongly shifted from the detection energy. The origin of this shift
remained unclear.  Moreover, no measurable Stokes shift was observed
in a recent paper Ref. \onlinecite{K2}, where the PL has been
measured together with the direct absorption by the layer of quantum
dots.

In this paper a simple theory of the PL from an array of quantum dots
is developed. We suggest that the process of photoexcitation of a dot
into its lowest optically-excited state does not contribute to the PL
signal. The proposed interpretation of the experimental data explains
large Stokes-like shift between the PL and PLE peaks as the average
distance between the two lowest optically-excited states. In what
follows we use the term ``ground state'' to denote the lowest
optically-excited state because it corresponds to the ground state of
the photoexcited carriers in the dot.

It is suggested that the PL and PLE lineshapes are completely
determined by the statistical distribution of the energy levels of
different quantum dots in the array. More precisely, it is determined
by the distribution of {\em pairs} of energy levels. It is shown that
such distribution is essentially two-variable. That is, there exists
a correlation in the positions of different energy levels in each
quantum dot throughout the array, however, such correlation is not
100\%. We show that this feature causes the difference in the
positions of the maxima of the PL and PLE spectra.

The fluctuations of energy levels due to the random potential caused
by the alloy composition fluctuations are studied in Section IV. We
show that the major part of the observed linewidth can be accounted
for by this mechanism. We also suggest that the first two excited
states observed in Ref.\ \onlinecite{P5} originate from the twofold
degenerate first excited level, when the degeneracy is lifted by the
random potential. The density of states and the two-level
distribution function, which accounts for the correlation in
energy-level positions, are calculated. The spectra determined by
these functions describe most of the features of the experimental
spectra.

Finally, the effects caused by fluctuations of the {\em shape} of the
quantum dots are discussed. It is suggested that the fluctuations of
the shape of the dots should increase the energy-level correlation.

\section{Origin of the Stokes-like Shift}

In the present paper we consider the case when both the electron and
the hole are confined in the quantum dot. Energy of the quantum dot
is measured from the unexcited state of the dot with no electrons and
holes. The term ``ground state'' refers to the ground state of the
quantum dot with an excited electron-hole pair, i.e.\ to the lowest
optically-excited state.

The key feature of an array of quantum dots that differs it from a
quantum well or bulk material is that there is no charge transfer
between the dots, or at least such transfer is strongly suppressed.
Different quantum dots therefore give independent additive
contributions in any optical experiment.\cite{C1}

The density of states for each quantum dot represents a set of
delta-function-like peaks, while the total density of states of the
whole system can be spread in a wide energy range due to the
inhomogeneous broadening. Several experiments have been reported
recently where the contributions to the luminescence from {\em
single} quantum dots were found.\cite{M1,K2,P6,A1} Single quantum
dots give extremely narrow sub-meV spectral lines. When, however, the
number of the excited quantum dots is large, a broad luminescence
peak is observed.\cite{P5,K2,P6}

We suggest that such a feature makes it difficult to probe optically
the ground states of the dots and thus causes an apparent Stokes-like
shift observed in Ref.\ \onlinecite{P5}. Probing of the ground states
of the dots may require special technique like high resolution or
time-resolved PL.

Indeed, let's consider the contribution to the PL from the process
when the photon is absorbed into the ground state of a dot and then
re-emitted. If the optical process is not phonon-assisted, the
energies of the incoming and outgoing photons are exactly equal. The
PL response can hardly be observed since it is hidden by the incident
beam scattered by other elements of the experimental environment.
This situation is quite different from that in quantum wells or bulk
semiconductors, where an electron excited to the conduction band has
always quantum states with smaller energies.  It can loose some
energy before it recombines in an {\em independent} optical process.

In a quantum dot, the emitted photon may have its energy below (or
above) that of the exciting light if the PL is phonon-assisted.
However, the phonon must be emitted (absorbed) together with photon
{\em in the same quantum process}. The probability of such process is
determined in higher-order perturbation theory in the electron-phonon
coupling constant and is therefore much smaller than the probability
of the direct transition.

In order to give a substantial contribution to the PL signal, the dot
must be pumped into one of its {\em excited} states. Therefore, the
minimum distance between the excitation and detection energies seen
in the spectra is equal to the distance between the ground and first
excited states of the dot. In fact, no Stokes shift was found between
the PL and the direct absorption by a layer of quantum dots measured
in Ref.\ \onlinecite{K2}.

\section{Positions of peaks in the PL and PLE spectra}

In this section the shape of the PL and PLE spectra is described
qualitatively. Different peaks observed in the spectra are assigned.
It is shown that existence of a random spread of interlevel
distances in quantum dots causes substantial deviations of the
positions of peaks seen in PLE and selectively-excited PL spectra.

We shall assume for simplicity that the energy relaxation in a
quantum dot occurs faster than recombination, so that the light is
always emitted from the ground state of the dot.

Let us first consider a simple model when the distances between
different energy levels are the same for all quantum dots, however,
there is a wide distribution of energy levels in the array. In other
words, let the picture of energy levels be the same in each quantum
dot but shifted randomly as a whole. This implies a 100\% correlation
in the positions of different energy levels in each dot.

If the conventional PL technique is used so that the pumping is
performed with energies well above the barriers between the dots,
all the dots are excited and emit light at their ground-state
energies. The emission then represents a broad peak due to the wide
distribution of the ground-state energies.

Let us now consider PLE and selectively-excited PL measurements. Here
only those dots are excited, which have one of their energy levels in
resonance with the exciting light. As suggested above, the dots
pumped into their ground states do not contribute to the spectra.
When the interlevel distances are the same in all dots, both PLE and
selectively-excited PL spectra would show a set of
delta-function-like peaks as sketched in the Fig.\ 1.

It is more convenient to start with the PLE (left). By fixing the
detection energy one selects the subset of all quantum dots in the
array with the ground-state energy $E_0=E_{detector}$ (we have
assumed that the light is always emitted from the ground state of the
dots).  The PLE signal appears when the excitation energy matches the
energy of an excited state ($E_{laser}=E_1$, $E_2$, etc.) in the
selected subset. Thus the first observed (lowest in energy) PLE peak
corresponds to the {\em first excited energy levels of such dots}
that have their ground-state level at the detection energy.

The analysis of the selectively-excited PL (right) is somewhat
more complicated but similar. The fixed energy of excitation selects
{\em several} subsets of all quantum dots such that $E_1=E_{laser}$,
$E_2=E_{laser}$, etc. A nonzero PL signal appears when the detection
energy matches the ground-state energy of one of the subsets. The
position of the first observed PL peak (highest in energy) thus
corresponds to the {\em ground state energy of such dots} that have
their {\em first excited} energy level at the excitation energy. The
distance between the first two PL peaks is equal to $E_{21}$, the
distance between {\em the first and the second} excited states of the
dots.

When interlevel distances are the same in all dots, all distances
between different peaks in the PL (PLE) series also remain the same,
while the whole picture shifts with the shift of the laser (detector)
energies. It means that positions of these peaks being plotted
against the laser (detector) energies must form a set of straight
lines with the unit slope.

The slopes of such lines obtained from the PLE and
selectively-excited PL spectra obtained in Ref.\ \onlinecite{P5} are
all less then 1. For the first PL and PLE peaks, e.g.,
$dE_{max,1}^{PL}/dE_{laser}=0.91$, $dE_{max,1}^{PLE}/
dE_{detector}=0.77$.\cite{P5}

It is easy to see that such deviation cannot be attributed to the
dependence of the interlevel distances on the ground-state energy.
Indeed, in this case the slopes obtained from the PL and PLE spectra
should be inverse of each other. We see, however, that
\begin{equation}
  \frac{dE_{max,m}^{PL}}{dE_{laser}}\neq
  \left(\frac{dE_{max,m}^{PLE}}{dE_{detector}}\right)^{-1}.
\end{equation}

We show below that existence of a spread of interlevel distances in
the array causes deviation of the positions of peaks
seen in the PL and PLE spectra in such a way that both derivatives in
Eq.\ (1) become less then 1.

To demonstrate this, let us assume that the interlevel distances in
the dots fluctuate randomly in some (narrow) energy interval. In this
case selection of a subset of quantum dots by fixing one of the
energy levels does not yet determine the positions of other energy
levels. Then one should observe a sequence of broadened peaks in both
PL and PLE.  The scale of broadening is determined by the
distribution of interlevel distances only, therefore it can be
narrower than the distribution of the absolute energies. As shown
below such behavior is natural if the spread is caused by a random
potential.

To understand how the {\em positions of maxima of the broadened
peaks} are shifted, it is convenient to draw a three-dimensional
picture shown in Fig.\ 2. Here the intensity measured by the detector
is plotted as a function of two variables --- the excitation and the
emission energies. Clearly, such a plot contains all information,
which both PL and PLE can provide. To get the shapes of the PL or PLE
spectra one simply has to slice the three-dimensional plot along the
$E_{laser}$ or $E_{detector}$ axis.

First, the intensity is zero in the half-plane
$E_{laser}<E_{detector}$. The intensity is nonzero only when
$E_{laser}$ and $E_{detector}$ are in resonance with the excited and
ground states of some dot {\em simultaneously}. If the spread of
interlevel distances is small, the three-dimensional plot has a shape
of a set of narrow {\em ridges}, elongated in the direction parallel
to the line $E_{laser}=E_{detector}$. Each ridge corresponds to the
optical process for which $E_{detector}=E_0$ and $E_{laser}=E_m$,
$m$=1, 2, etc. The width of each ridge is determined by the spread of
the corresponding interlevel distances, while the length is larger
and represents the large spread of the ground-state energies. The
distance between the ridge and the line $E_{laser}=E_{detector}$ is
determined by the average distance between the ground and the
corresponding excited states of the dots.

The inset in Fig.\ 2 shows a fragment of the same plot in the isoline
projection (dashed lines). Crosses and pluses show the positions of
the maxima in the PL and PLE crossections respectively. The ``PL and
PLE lines'' are the loci of the maxima observed in the
PL and PLE spectra.

As one can see, the positions of the maxima deviate from the major
axis of the ridge and from each other. Indeed, these positions lie in
such points where the crossection line (parallel to one of the axes)
is tangential to the lines of equal intensity. Note that the
deviation of the positions of the maxima is such that both
derivatives in the Eq.\ (1) are less than 1.

Both lines intersect exactly at the maximum of the ridge, which
coordinates give the average positions of the ground and the excited
states of all the dots in the array.

In general, if the oscillator strengths are the same for all allowed
transitions, the intensity as a function of the excitation and
detection energies is proportional to the mutual level-level
distribution function: $I(E_{detector},E_{laser})\propto\sum_m
P(E_{detector}= E_0,E_{laser}=E_m)$, where $E_m$ are the energy-level
positions, $E_0$ being the position of the ground state. In the
following section we derive the shape of the distribution function
$P(E_0,E_m)$ when the spread of the energy-level positions is
caused by the composition fluctuations. We show that a quite
complicated PL and PLE lineshape observed in Ref.\ \onlinecite{P5}
can be sufficiently well described in this way.

\section{fluctuations of energy levels in quantum dots}

In this section we study the properties of the statistical
distribution of the energy levels in quantum dots caused by a
white-noise random potential. We present a strong evidence that the
major part of the observed spread of the PL and PLE peaks is
caused by universal composition fluctuations in the dots. These
fluctuations are a generic property of semiconductor alloys and
produce a theoretical limit for unification of quantum dots in an
array.

We also suggest that the first two excited levels that reveal
themselves in the PLE and selectively-excited PL experiments
originate {\em from a single doubly degenerate level} when the
degeneracy is lifted by the random potential.

In semiconductor alloys the lattice sites are occupied randomly with
two types of substitutional atoms. We ignore here the correlation
between occupation of different sites. The composition $x$ averaged
over a small volume always fluctuates. The order of magnitude of the
fluctuations is inversely proportional to the square root of the
volume over which the averaging is performed. Though usually small,
this effect can be important in extremely small structures.

In small quantum dots such composition fluctuations cause shifts of
energy levels from the positions determined by the average
composition $x_0$. The ``local'' composition $x$ varies from dot to
dot, and also {\em inside} the dot. According to that, the energy
levels fluctuate from dot to dot. Shifts of the energy levels in
different quantum dots are independent from each other. Inside a
single quantum dot the shifts of different energy levels are
correlated. Note, however, that such correlation is not 100\% (as it
would be if there was only one fluctuating parameter like the
diameter of a dot).

In order to calculate the statistical distribution of energy levels
caused by composition fluctuations we use the method developed by
Efros and Raikh (for a review see Ref.\ \onlinecite{RE}). This method
is applicable if the size of the wave function is much larger than
the lattice constant. In this case one needs to know only the shape
of the wave function in a quantum dot and the slope of the dependence
of the gap on composition, $dE_g/dx$, at average composition $x_0$.

The result depends on whether the unperturbed energy levels are
degenerate or not. Without degeneracy the distribution of energy
levels $E_m$ is Gaussian with the standard deviation given by
\begin{equation}
\sigma_m^2=\overline{\epsilon_m^2}=
  \gamma\int d^3r\ \psi_m^4({\bf r}),
\end{equation}
where $\epsilon_m=E_m-\overline E_m$, $\psi_m$ is the wave function
(real) corresponding to the energy level $E_m$, and $\gamma$ is given
by
\begin{equation}
\gamma=\left(\frac{dE_g}{dx}\right)^2\frac{x_0(1-x_0)}{N},
\end{equation}
where $N$ is the number of lattice sites per unit volume.

The {\em covariance} between $\epsilon_m$ and $\epsilon_n$ can be
obtained in a similar way:
\begin{equation}
\overline{\epsilon_m\epsilon_n}=\rho\sigma_m\sigma_n
  =\gamma\int d^3r\ \psi_m^2({\bf r})\psi_n^2({\bf r}).
\end{equation}
The coefficient $\rho$, $\rho\leq1$, is called the coefficient of
correlation between $E_m$ and $E_n$.

To find the shape of the PLE and selectively-excited PL spectra we
need the mutual distribution function for $E_0$ and $E_m$. The most
general form of the two-variable Gaussian distribution is given by
\begin{eqnarray}
&G&_2(\epsilon_0,\epsilon_m;\sigma_0,\sigma_m,\rho)=
  \frac{1}{2\pi\sigma_0\sigma_m\sqrt{1-\rho^2}}\nonumber\\
&&\times\exp\left\{-\frac{1}{2(1-\rho^2)}
  \left[\frac{\epsilon_0^2}{\sigma_0^2}
    -2\rho\frac{\epsilon_0\epsilon_m}{\sigma_0\sigma_m}
    +\frac{\epsilon_m^2}{\sigma_m^2}\right]\right\},
\end{eqnarray}
This is just an analytical expression for the shape of the
ridge, discussed in the previous section. The ridge is strongly
elongated when the parameter $\rho$ is close to 1. It is equal to 1 in
the limiting case when an exact relation between the energy-level
positions exists, so that the two-variable statistical distribution
becomes effectively one-variable. The ratio $\sigma_m/\sigma_0$
determines the orientation of the ridge. The ridge is parallel to the
line $E_{laser}=E_{detector}$ when $\sigma_m=\sigma_0$.

The situation is, however, more complicated if a degeneracy exists.
The random potential shifts the degenerate energy level and lifts the
degeneracy. The distribution of energies in the vicinity of an
unperturbed degenerate level appears to be not Gaussian. The mutual
two-variable distribution function for a transition between $E_0$ and
$E_m$ is not Gaussian either. Instead, it has a shape of two close
parallel ridges corresponding to each of the split-off energy levels.

For simplicity we restrict ourselves to the axially symmetric quantum
dots, where each energy level except the ground state is doubly
degenerate. In this case it appears to be possible to derive the
general form of the distribution function without knowledge of the
shape of the wave functions in the quantum dot.

For an axially symmetric system two degenerate wave functions with
the angular momentum $|m|$ have the form $\psi_{m\pm}({\bf
r})=\psi_m(r) e^{\pm im\phi}$, where $\psi_m$ can be made real. The
positions of energy levels $E_{m\pm}$, split and shifted by random
potential of the particular configuration, can be obtained as the
eigenvalues of the secular matrix
\begin{equation}
\delta H=\left[
\begin{tabular}{cc}
$u$ & $x+iy$\\
$x-iy$ & $u$
\end{tabular}
\right],
\end{equation}
where the matrix elements $u$ and $x+iy$ take random values in each
quantum dot and are given by:
\begin{mathletters}
\begin{eqnarray}
u&=&\sqrt{\gamma}\int d^3r\ V({\bf r})\psi_m^2({\bf r}),\\
x+iy&=&\sqrt{\gamma}\int d^3r\ V({\bf r})\psi_m^2({\bf r})
   e^{2im\phi}.
\end{eqnarray}
\end{mathletters}
Here $V({\bf r})$ is the white-noise random potential with correlator
$<V({\bf r})V({\bf r}')>=\gamma\delta({\bf r}-{\bf r}')$.

Eigenvalues of $\delta H$ are $\epsilon_\pm=u\pm\sqrt{x^2+y^2}$
(the energy is measured from the unperturbed energy level). It is
easy to see that $u$, $x$, and $y$ are {\em independent} Gaussian
random variables with $\overline{u^2}=\sigma_m^2$,
$\overline{x^2}=\overline{y^2}=\sigma_m^2/2$, $\sigma_m^2$ being
given by Eq.\ (2).

The density of states in the vicinity of $E_m$ (the distribution
function for $\epsilon_m=E_m-\overline E_m$) is given by
\begin{eqnarray}
P(\epsilon_m)&=&\sum_\pm\int\!\int\!\int
  dudxdy\ \delta(\epsilon_m-u\mp\sqrt{x^2+y^2})\nonumber\\
  &&\times G_1(u;\sigma_m)
  G_1(x;\frac{\sigma_m}{\sqrt{2}})G_1(y;\frac{\sigma_m}{\sqrt{2}})
  \nonumber\\
&=&D_0(\epsilon_m;\sigma_m),
\end{eqnarray}
where $G_1$ is the standard one-variable Gaussian distribution,
$G_1(\epsilon;\sigma)\equiv\sigma^{-1}G_1(\epsilon/\sigma;1)$,
$G_1(z;1)=(2\pi)^{-1/2}\times$ $\exp(-z^2/2)$; and we have
introduced the notation for the distribution function
$D_0(\epsilon;\sigma)\equiv\sigma^{-1}D_0(\epsilon/\sigma;1)$,
\begin{equation}
D_0(z;1)=\frac{2}{3}\sqrt{\frac{2}{\pi}}e^{-z^2/2}+
  \frac{2z}{3\sqrt{3}}e^{-z^2/3}\text{erf}(\frac{z}{\sqrt{6}}).
\end{equation}
The meaning of the index 0 in $D_0(\epsilon;\sigma)$ will soon become
clear.

The Eq.\ (9) describes a bell-shaped curve, which determines the
density of states and, hence,  the {\em absorption spectrum} in the
vicinity of the excited level $E_m$. It's shape can be fairly well
approximated by a Gaussian with the effective dispersion
$\sigma_{eff}=\sigma_m\sqrt{2}$, as shown in Fig.\ 3.

In order to find the mutual two-level distribution function that
determines the shape of the PL and PLE spectra it is important to
account for the correlation between matrix elements of $\delta H$ and
the shift of the ground-state $\epsilon_0$.

It is useful to note that the expression for the matrix element $u$ is
the same as the expression for the shift of a non-degenerate level
with the wave function $\psi_m(r)$. The value of $u$ is, therefore,
{\em correlated} with $\epsilon_0$ via the same two-variable Gaussian
distribution as in Eq.\ (5):
$P(\epsilon_0,u)=G_2(\epsilon_0,u;\sigma_0,\sigma_m,\rho)$, with the
parameters $\sigma_0$, $\sigma_m$, and $\rho$ given by Eqs.\ (2),
(4). The parameters $x$ and $y$ are statistically independent of
$\epsilon_0$ and $u$, because the corresponding covariances become
zero when the integration over the azimuthal angle $\phi$ is
performed. Thus, we obtain:

\begin{eqnarray}
P(\epsilon_0,\epsilon_m)&=&\sum_\pm\int\!\int\!\int
  dudxdy\ \delta(\epsilon_m-u\mp\sqrt{x^2+y^2})\nonumber\\
  &&\times G_2(\epsilon_0,u;\sigma_0,\sigma_m,\rho)
  G_1(x;\frac{\sigma_m}{\sqrt{2}})G_1(y;\frac{\sigma_m}{\sqrt{2}})
  \nonumber\\
&=&G_1(\epsilon_0;\sigma_0)
  D_\rho(\epsilon_m-\epsilon_0\rho\frac{\sigma_m}{\sigma_0};\sigma_m),
\end{eqnarray}
where the function
$D_\rho(\epsilon;\sigma)\equiv\sigma^{-1}D_\rho(\epsilon/\sigma;1)$,
and

\begin{eqnarray}
D_\rho(z;1)&=&\frac{2\sqrt{\mu-1}}{\sqrt{\pi}\mu}
  \exp\left(-\frac{z^2}{\mu-1}\right)\nonumber\\
&+&\frac{2z}{\mu^{3/2}}\exp\left(-\frac{z^2}{\mu}\right)
  \text{erf}\left[\frac{z}{\sqrt{\mu(\mu-1)}}\right].
\end{eqnarray}
Here $\mu=3-2\rho^2$. The function $D_0(\epsilon;\sigma)$ defined
by Eq.\ (9) is a particular case of (11) with $\rho=0$.

The function $D_\rho(\epsilon_m;\sigma_m)$ gives the distribution of
the energy sublevels in the vicinity of $\overline E_m$, when {\em
the position of the ground state is fixed}. It is normalized in such
a way that $\int d\epsilon\ D_\rho(\epsilon;\sigma)=2$ (according to
the twofold degeneracy), and $\int d\epsilon\ \epsilon^2
D_\rho(\epsilon;\sigma)=2\sigma^2(2-\rho^2)$. The function
$D_\rho(\epsilon;\sigma)$ is symmetric in $\epsilon$. It has one
maximum when $\rho\leq1/\sqrt{2}$ and two maxima, when
$\rho>1/\sqrt{2}$. The maxima become more pronounced when $\rho$
tends to 1.

The function $P(\epsilon_0,\epsilon_m)$ determined by Eq.\ (10)
gives the probability density for a quantum dot to have the ground
state at the energy $\overline E_0+\epsilon_0$ and an excited state
at the energy $\overline E_m+\epsilon_m$. It is proportional to the
intensity measured by the detector at fixed excitation and detection
energies. Hence, it determines the shape of both PL and PLE spectra
when the proper argument is fixed.

The function $P(\epsilon_0,\epsilon_m)$ appears to be not very
sensitive to the ratio $\sigma_m/\sigma_0$. It is, however, quite
sensitive to the value of the correlation coefficient $\rho$.
This function is plotted in Fig.\ 4 for $\sigma_m=\sigma_0$ and two
different values of $\rho$. As shown in the next section, it is
natural for the coefficient $\rho$ to be close to 1. The value
$\rho=0.94$ [Fig.\ 4(a)] gives the best fit to the experimental data.
When $\rho$ tends to 1 [Fig.\ 4(b)] the intensity in the dip between
two maxima approaches zero. The lineshapes of the spectra described
by Eq.\ (10) are discussed in detail in the next section.

\section{discussion}

We replot the positions of the PL and PLE maxima observed in Ref.\
\onlinecite{P5} on the combined plot in Fig.\ 5 to illustrate that
the experimental behavior is in agreement with our consideration
(compare with the inset in Fig.\ 2). The dotted line shows unit
slope, $E_{laser}=E_{detector}$. Three lines correspond to the
positions of two maxima observed in each PL curve (crosses) and one
--- in each PLE curve (pluses). Positions of only one PLE maximum for
each $E_{detector}$ are shown because the second PLE maximum is seen
not clear enough. The constant value of 1290 meV has been subtracted
from all energies.

Three lines shown in Fig.\ 5 are consistent with the qualitative
picture of two close parallel ridges. As suggested above, these
ridges correspond to the two optical processes where
$E_{detector}=E_0$ and $E_{laser}=E_{1\pm}$; $E_0$ being the
ground-state energy and $E_{1\pm}$ --- the energies of the first and
second excited states.

We use the notation $E_{1\pm}$ instead of $E_{1,2}$ for the two
excited states according to the idea that they originate from the
twofold-degenerate first excited energy level $E_1$ when the
degeneracy is lifted by random potential. This idea is supported by
the fact that the ratio of interlevel distances appears to be
$(\overline E_{1+}-\overline E_{1-})/(\overline E_{1-}-\overline
E_0)\approx 0.6$. For a cylindrical quantum well with infinite
potential walls the corresponding ratio is about $(\overline
E_{21}/\overline E_{10})=1.3$.

The point of intersection of two lines gives the position of the
maximum of the first ridge. The difference of 55 meV in its
coordinates is nothing but the observed Stokes-like shift between
emission and absorption energies. The distance between two ridges
measured along the $E_{laser}$ axis gives $\overline E_{1+}-\overline
E_{1-}$, the mean splitting of the excited state $E_1$.

Using Fig.\ 5 we may conclude that the average distance between the
ground and (split) first excited state is about 75 meV, while the
average splitting of the excited state is approximately 40 meV.

Fig.\ 6 shows the comparison between the presented theory and the
experiment. The experimental spectra from Ref.\ \onlinecite{P5} are
replotted in Fig.\ 6(a). The theoretical PLE and selectively-excited PL
spectra [Fig.\ 6(b)] are obtained with the use of Eq.\ (10) by fixing
$\epsilon_0$ or $\epsilon_m$ respectively. Experimental values for
the average positions of the energy levels $\overline E_0=1276$ meV
and $\overline E_1=1351$ meV are used.

The Fig.\ 6(b) is essentially the Fig.\ 4(a), replotted in the wavelength
scale for better comparison. The value $\rho=0.94$ gives the best fit
for the data from Ref.\ \onlinecite{P5}. It is seen that the curves
obtained reproduce all the features of both PL and PLE spectra.

The structure of the expression (10) is such that if $\epsilon_0$
is fixed, it describes a curve symmetric around the point
$\epsilon_m=\epsilon_0\rho\sigma_m/\sigma_0$. Thus (if
$\rho>1/\sqrt{2}$) the PLE line should reveal two {\em symmetric}
peaks at any detection energy. This is exactly the behavior seen in
the experiment. When the detection energy is changed, the whole
spectrum must shift linearly with it. Indeed, Fig.\ 5 shows that the
positions of PLE maxima depend linearly on the detection energy.
The slope of the PLE curve in Fig.\ 5 gives the ratio
$\sigma_1/\sigma_0$, which appears to be close to 1.

If the position of the excited level, $\epsilon_m$, is fixed, the
lineshape is asymmetric. When $\rho$ is close to 1, the PL lineshape
also shows two maxima, however, there is a peculiar interplay in their
magnitudes when the excitation energy is changed. This also matches
the experimental data quite well.

Such interplay can be easily understood. First (second) PL maximum
corresponds to the distribution of ground-state energies of such
subset of quantum dots, for which the lowest (highest) split-off
level coincides with $E_{laser}$. When $E_{laser}$ is below $\overline
E_m$, the amounts of first type of dots is larger than that of the
second type. When $E_{laser}$ is larger than $\overline E_m$, the dots
of the second type win.

The interplay is absent in PLE, because for each fixed ground-state
energy there is always equal amount of lowest and highest split-off
levels.

The magnitudes of $\sigma_1$ and $\sigma_0$ can be obtained
independently as follows. For $\rho=0.94$, the maximum of
$D_\rho(\epsilon;\sigma)$ lies at $\epsilon=0.765\sigma$. Then, from
the position of the PLE maximum, $\sigma_1$=14.5 meV/0.765=19 meV.
The spread of the ground states $\sigma_0$ can be determined
independently from the nonselective PL [curve 9 in Fig.\ 6(a)]. It
gives $\sigma_0$=18.2 meV.

If we know the shape of the wave function, we may find the values for
the parameters $\sigma_{0,1}$ and $\rho$, using Eqs.\ (2,4). As a
guess, we may try the wave functions for the cylindrical quantum well
with infinite walls:

\begin{equation}
\psi_m({\bf r})\propto\cos(\frac{\pi r_\bot}{h})
  J_m(\frac{\nu_m r_\|}{R}),
\end{equation}
where $h$ and $R$ are correspondingly the thickness and the radius of
the quantum dot, and $\nu_m$ is the root of the Bessel function
$J_m$. The integrals of interest, $J_{mn}=\int d^3r\
\psi_m^2\psi_n^2$, are equal to: $J_{00}=2.098$, $J_{11}=1.552$, and
$J_{01}=1.435$.

It is easy to find all parameters in this approximation. First, let
us estimate the magnitude of the spread. To find $\sigma_0$ one has
to know the volume of the quantum dot. Taking it to be the volume of
the cylinder with the thickness 2.5 nm and diameter 25 nm,\cite{P5}
and using the parameters of InGaAs: $x_0=0.5$, lattice constant
$a=0.585$ nm and $(dE_g/dx)=1.16$ eV (both at $x=0.5$),\cite{LB} we
obtain $\sigma_0=13$ meV. This value is less than the experimental
value of 18.2 meV. It, however, shows that at least a significant
part of the spread is caused by the composition fluctuations.
There are, of course, some other reasons for spreading. The total
spread, however, cannot be less than the calculated value.

Two remaining dimensionless parameters are: $\rho=0.795$, and
$(\sigma_0/\sigma_1)=1.16$. Though the ratio $\sigma_0/\sigma_1$ is
in a reasonable agreement with the experiment, the value of the
correlation coefficient $\rho$ is significantly less than the
experimental value of 0.94. Note that the dimensionless parameters
depend only on the form of the wave functions. Though it is possible
to relate the discrepancy in $\rho$ to the unknown shape of the real
wave function, there is a more serious reason for this coefficient to
be closer to 1.

Among the other causes of spreading of energy levels, which are
not taken into account in the presented theory, there are {\em
fluctuations of the shape} of quantum dots. The distortion of the
shape of a quantum dot, even when small, cannot be represented as a
potential perturbation in the Schr\"odinger equation. It's effect on
the positions and splitting of the energy level $E_m$ can, however,
be described by an effective secular matrix $\delta H$ of the same
form as Eq.\ (6). The only difference is that the matrix elements $u$
and $x+iy$ are not described by the Eqs.\ (7) any more. Instead, they
are determined by the integrals of the derivative of the unperturbed
wave function at the boundary of the quantum dot.\cite{ED}

Thus, the effect of the shape fluctuations is only in renormalizing
the parameters $\sigma_0$, $\sigma_m$, and $\rho$. It is important,
however, that for the case of pure shape fluctuations, the parameter
$\rho$, defined as a correlation coefficient between $u$ and
$\epsilon_0$, is exactly equal to 1. The reason is that in an
axially-symmetic quantum dot, the normal derivative of the wave
function at the boundary is just a number rather than a function of
coordinate. Therefore, there should be an exact relation between
$u$ and $\epsilon_0$ {\em for each particular shape distortion}.
It is natural to assume that the effective value of $\rho$ is closer
to 1 when both mechanisms are involved, than it is in the case of
pure composition fluctuations.

\section{Conclusion}
In the present paper a simple theory is developed, which allows to
describe selective photoluminescence data from an array of quantum
dots with random parameters. The theory explains large apparent
Stokes-like shift between emission and absorption energies as the
average distance between the ground and first excited energy levels
in the dots.

It is shown that existence of a random spread of the interlevel
distances in the dots causes deviation of the positions of the maxima
of the peaks seen in the PL and PLE spectra. Such deviation can make
it difficult to determine the properties of the statistical
distribution of energy levels in the array. It is suggested how the
proper parameters of the statistical distribution may be obtained
from the experimental data.

The random shifts and splittings of energy levels caused by a
white-noise random potential in the dots are studied. The density of
states and mutual level-level distribution function are obtained for
the case of axially symmetric quantum dot. The energy-level
distribution and the resulting PLE and selectively-excited PL spectra
appear to be close to that observed in the experiment (see Fig.\ 6).
It is shown that the major part of the spread of energies observed in
the experiment originates from the random potential caused by the
composition fluctuations. It is also suggested that the random
fluctuations of the shape of the dots also contribute to the spread.

\section*{Acknowledgments}
I am grateful to A. L. Efros for formulating the problem and for
numerous illuminating discussions. I would like to thank P. M.
Petroff for providing the experimental data prior to the publication.
I appreciate important comment by M. E. Raikh and useful discussions
with P. M. Petroff, E. I. Rashba, J. M. Worlock, and F. G. Pikus.
This work was supported by the Center for Quantized Electronic
Structures (QUEST) of UCSB under subagreement KK3017.

\begin{figure}
\caption{A schematic diagram of the PLE and selectively-excited PL
intensities when interlevel distances in all quantum dots are the
same. Each curve is the spectrum taken at fixed excitation energy
(PL) or fixed detection energy (PLE), which are denoted by pluses.}
\end{figure}

\begin{figure}
\caption{A schematic shape of the intensity as a function of the
excitation and detection energies for the system with two excited
states. The left ``ridge'' corresponds to the absorption by the first
excited states of the dots. The ``PL'' and ``PLE lines'' in the inset
are the loci of the positions of the maxima of intensity in the PL
and PLE crossections respectively. The dashed lines show the lines of
equal intensity. The maxima in a crossection occur at such points
where the crossection lines (horizontal and vertical solid lines) are
tangential to the lines of equal intensity.}
\end{figure}

\begin{figure}
\caption{The function $D_0(z;1)$ (full line), as given by Eq.\ (9),
determines the density of states in the vicinity of the excited
level. The approximation by a Gaussian with an effective dispersion
(dashed line) is also plotted.}
\end{figure}

\begin{figure}
\caption{The two-level mutual distribution function
$P(\epsilon_0,\epsilon_1)$ given by Eq.\ (10) is plotted for fixed
$\epsilon_0=0$ (the ``PLE'' curve) and for different fixed
$\epsilon_1$ (the set of ``PL'' curves). The curves are shifted
upward arbitrarily to ease the reading. The correlation coefficient
is: (a) $\rho=0.94$ and (b) $\rho=1$.}
\end{figure}

\begin{figure}
\caption{The positions of the PL (crosses) and PLE (pluses) maxima
taken from the spectra in Ref.\ 1 are shown in the combined plot.
The resulting crossing lines make clear how the positions of maxima
deviate from each other and from the major axis of the first
``ridge'' (dashed line). The crossection point is the maximum of the
ridge. The second set of the PLE maxima is not shown because
the corresponding peaks are seen not very clear in the spectra. A
constant value of 1290 meV has been subtracted from all energies. The
dotted line shows the unit slope, $E_{laser}=E_{detector}$.}
\end{figure}

\begin{figure}
\caption{(a) The experimental data from Ref.\ 1. replotted for
comparison. The upper curve is the PLE spectrum obtained with the
detection energy marked by an arrow. Pluses above the PLE spectrum
show the fixed excitation energies for each of the PL spectra. The
curve 9 corresponds to the above-barrier excitation energy. (b) The
PLE and selectively-excited PL lineshapes obtained from Eq.\ (10)
with parameters $\rho=0.94$, $\sigma_0=\sigma_1=19$ meV. Average
energy positions of the ground and first excited energy levels are
$\overline E_0=1276$ meV, and $\overline E_1=1351$ meV.}
\end{figure}


\begin{references}
\bibitem[*]{RR} e-mail: rr@physics.utah.edu.
\bibitem{P5} S. Fafard, D. Leonard, J. L. Merz, and P. M. Petroff,
  \apl {\bf 65}, 1388, 1994.
\bibitem{M1} J.-Y. Marzin, J.-M. G\'erard, A. Izra\"el, G. Bastard,
  D. Barrier, \prl {\bf 73}, 716, 1994.
\bibitem{K2} M. Grundmann, J. Christen, N. N. Ledentsov, J. B\"ohrer,
  D. Bimberg, S. S. Ruvimov, P. Werner, U. Richter, U. G\"osele,
  J. Heydenreich, V. M. Ustinov, A. Yu. Egorov, A. E. Zhukov,
  P. S. Kop'ev, Zh. I. Alferov, \prl {\bf 74}, 4043, 1995.
\bibitem{P6} S. Fafard, R. Leon, D. Leonard, J. L. Merz,
  P. M. Petroff, \prb {\bf 50}, 8086, 1994.

\bibitem{C1}In fact, in order to consider different quantum dots
independently it is necessary that average distance between two
resonant quantum dots be larger than the wavelength of light. Though
average distance between the dots may be smaller than the light
wavelength (as it is the case in Ref.\ \onlinecite{P5}), the energy
levels of close quantum dots coincide seldomly. If, however, the
average distance between resonant quantum dots is smaller than the
wavelength, the whole array may exhibit some collective-mode effects.

\bibitem{A1} A. Zrenner, L. V. Butov, M. Hagn, G. Abstreiter,
  G. B\"ohm, G. Weimann, \prl {\bf 72}, 3382, 1994.
\bibitem{RE} A.L. Efros and M.E. Raikh, {\em Effect of Composition
  Disorder on the Electronic Properties of Semiconducting Mixed
  Crystals}, in: {\em Optical Properties of Mixed Crystals} ed. by R.J.
  Elliot   and I.P. Ipatova, Elsevier, 1988.
\bibitem{LB}Landolt-B\"ornstein: {\em Numerical Data and Functional
  Relationships in Science and Technology}, edited by O. Madelung
  (Springer-Verlag, Berlin, 1982), Vol. 17, Pts. a and b.
\bibitem{ED} See, e.g., C. C. Johnson, {\em Field and wave
electrodynamics}, McGraw-Hill, N.Y., 1965.
\end{references}
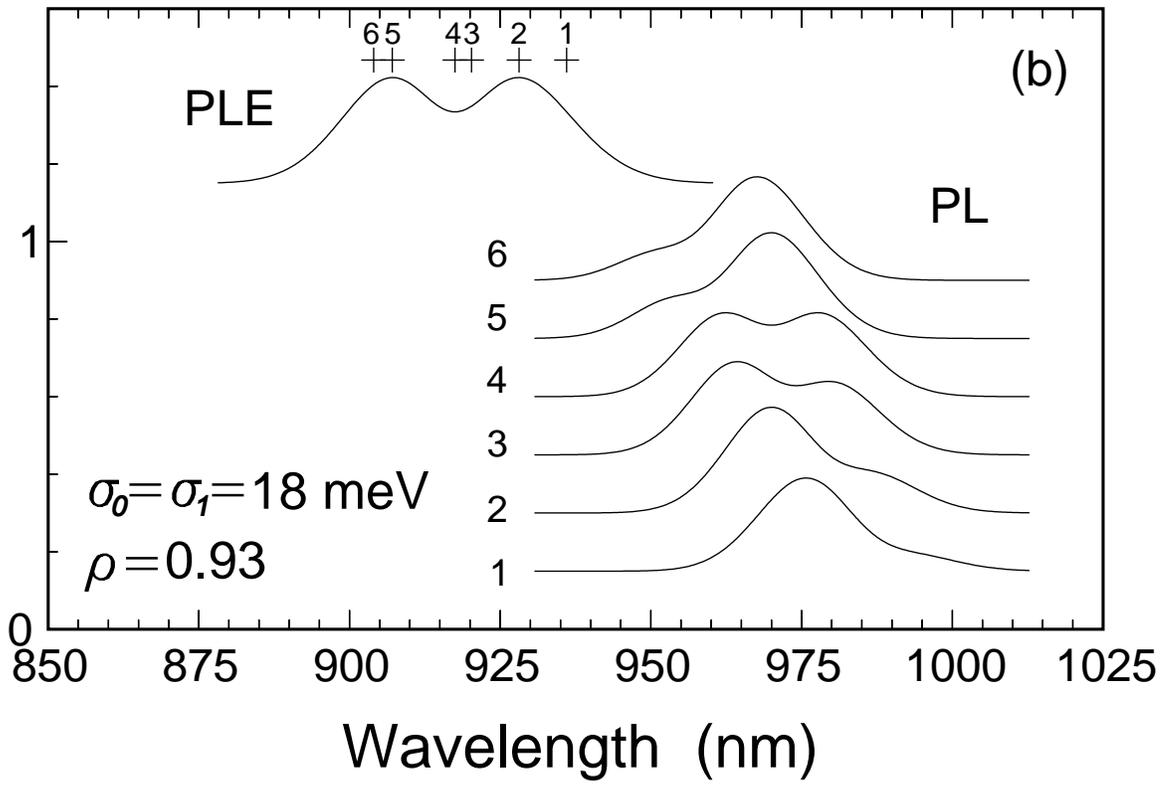
\end{document}